\documentclass[aps,prl,twocolumn,showpacs]{revtex4-1}
\usepackage[utf8]{inputenc}
\usepackage{graphicx}
\usepackage{amsmath}
\usepackage{hyperref}
\usepackage{units}
\usepackage{amsmath, amssymb}
\usepackage{natbib}
\usepackage{amsfonts}
\usepackage{textcomp}
\usepackage{verbatim}
\usepackage{natbib}
\usepackage{color}

\begin{document}

\title{Roton-type mode softening in a quantum gas with cavity-mediated long-range interactions}
\author{R.~Mottl}
\author{F.~Brennecke}
\author{K.~Baumann}
\author{R.~Landig}
\author{T.~Donner}
\author{T.~Esslinger}
\email{esslinger@phys.ethz.ch}
\affiliation{Institute for Quantum Electronics, ETH Z\"{u}rich,
8093 Z\"{u}rich, Switzerland}

\begin{abstract}
Long-range interactions in quantum gases are predicted to give
rise to an excitation spectrum of roton character, similar to that
observed in superfluid helium. We investigate the excitation
spectrum of a Bose-Einstein condensate with cavity-mediated
long-range interactions, which couple all particles to each other.
Increasing the strength of the interaction leads to a softening of
an excitation mode at a finite momentum, preceding a superfluid to
supersolid phase transition. We study the mode softening
spectroscopically across the phase transition using a variant of
Bragg spectroscopy. The measured spectrum is in very good
agreement with ab initio calculations and, at the phase
transition, a diverging susceptibility is observed. The work paves
the way towards quantum simulation of long-range interacting
many-body systems.
\end{abstract}
 \maketitle

Experiments with ultracold gases have succeeded in realizing a
variety of quantum many-body phases by exploiting the tunability
of short-range atom-atom interactions \cite{bloch2008}. Creating
quantum gases with long-range interactions
\cite{griesmaier2005,ni2008,deiglmayr2008,lu2011} is motivated by
the prospect of observing previously unexplored phenomena and
phases \cite{lahaye2009}. In particular, long-range interactions
in a Bose-Einstein condensate (BEC) have been predicted
\cite{santos2003,o'dell2003,cherng2009,henkel2010} to give rise to
an excitation spectrum similar to the roton spectrum observed in
superfluid helium \cite{yarnell1958}. A roton spectrum emerges
from density-density correlations which can be induced by
short-range van-der-Waals interactions as in liquid helium or by
momentum-dependent long-range interactions in dilute quantum gases
\cite{santos2003}. Such a mode softening at finite momentum has
been proposed as a possible route to a supersolid phase
\cite{pomeau1994}.

Important steps towards realizing quantum gases with long-range
interactions have been achieved by cooling atomic species with
large magnetic dipole moment to quantum degeneracy
\cite{griesmaier2005,lu2011}  and by the efforts to prepare
ultracold polar molecules in their ground state
\cite{ni2008,deiglmayr2008}. Effects of long-range interactions in
quantum gases have been observed in the anisotropic expansion, in
collective as well as Bloch oscillations, and in the stability of
interacting dipolar BEC
\cite{lahaye2007,fattori2008,bismut2010,koch2008}. However, a
roton-type mode softening requires very strong long-range
interactions and is challenging to observe in quantum gases.

A different approach to long-range interactions in cold gases
makes use of the radiative coupling between electric dipoles
induced by off-resonant laser light, which has been shown
theoretically to cause a roton-like minimum in the energy spectrum
of a BEC \cite{o'dell2003,henkel2010}.  A large enhancement of
such radiative coupling is achieved by placing the dipoles into an
optical resonator, which leads to strong global atom-atom
interactions
\cite{munstermann2000,asboth2004,maschler2005,slama2008a}.

%%%%%%%%%%%%
\begin{figure}
\centering
\includegraphics[width = \columnwidth]{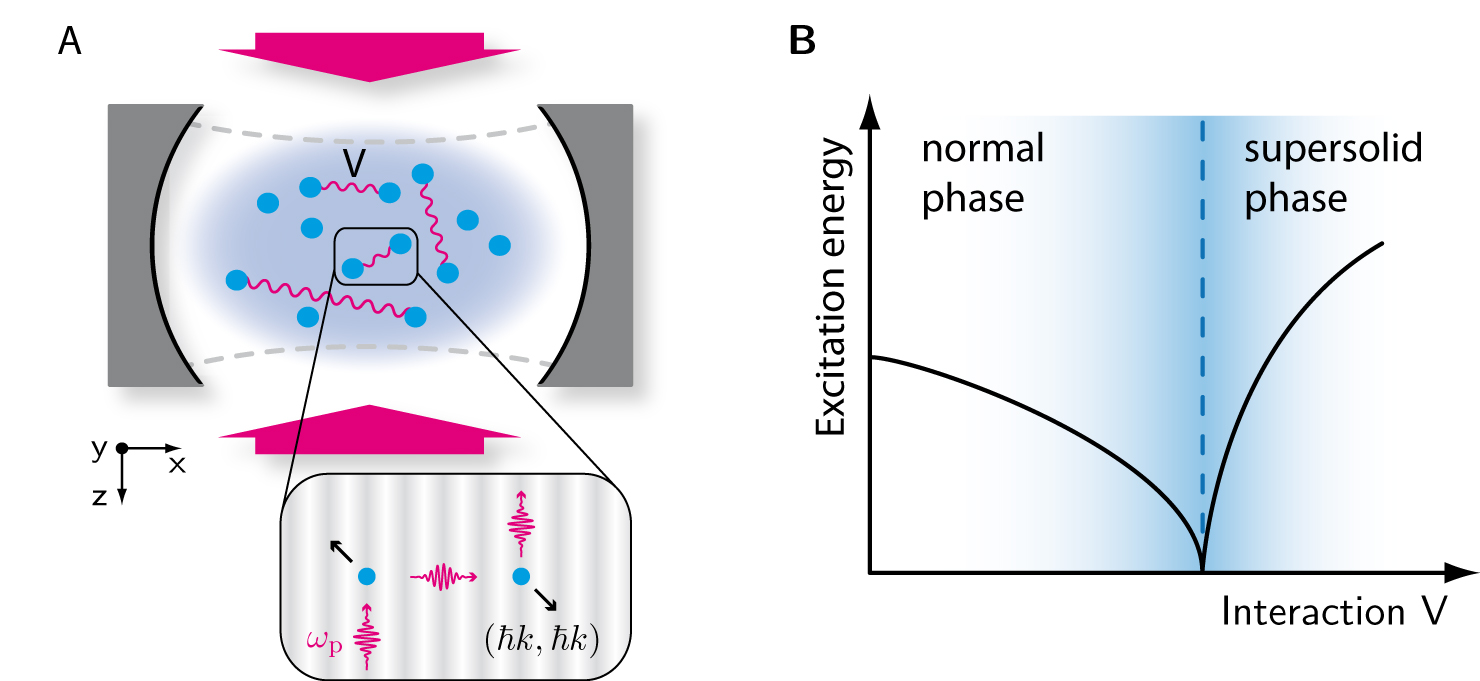}
\caption{
Experimental scheme and mode softening. (\textbf{A}) A $^{87}$Rb
BEC inside a Fabry-Perot resonator is transversally illuminated by
a far red-detuned standing-wave laser field. In a quantized
picture, atoms off-resonantly scatter photons from the pump field
into a close-detuned TEM$_{00}$ cavity mode and back at rate $N
V/\hbar$ (see text), creating and annihilating pairs of atoms in
the superposition of momenta $(p_x,p_z )=(\pm \hbar k, \pm \hbar
k)$ (see close-up displaying one of four possible processes). This
results in global interactions between all $N$ atoms. The
interaction strength $V$ is controlled via the power of the
transverse laser field. (\textbf{B}) The cavity-mediated atom-atom
interaction causes a softening of a collective excitation mode at
momenta $(\pm \hbar k, \pm \hbar k)$, and a diverging
susceptibility (blue shade) at a critical interaction strength
(dashed line).
}\label{figure1}
\end{figure}
%%%%%%%%%%%%%

We create long-range interactions in a BEC by placing it into an
optical high-finesse cavity and irradiating the atoms with a
transverse pump laser, which is far detuned from the atomic
transition frequency \cite{domokos2002,nagy2008,baumann2010}
(Fig.~1A). The induced electric dipoles of the atoms oscillate at
the pump laser frequency and collectively couple to a single
cavity mode, which is detuned from the pump frequency by a few
cavity linewidths. In turn, the induced cavity field leads to an
ac Stark shift in the atoms. The resulting atom-atom interaction
extends over the entire atomic cloud and is tunable in strength.
For a critical strength of the long-range interactions, the BEC
undergoes a quantum phase transition to a supersolid phase with
checkerboard density order \cite{nagy2008}, as observed in
\cite{baumann2010}. At the phase transition a discrete spatial
symmetry provided by the cavity mode structure is broken, thus
establishing nontrivial diagonal long-range order. We have
developed a method to measure the excitation spectrum across this
phase transition. The method combines Bragg spectroscopy
\cite{stenger1999} and cavity-enhanced Bragg scattering, and we
use it to observe a notable change in the dispersion relation
while crossing the phase transition. This is complemented by a
diverging susceptibility, in agreement with the second-order
nature of the phase transition (Fig.~1B).

The cavity-mediated long-range interaction is described by the
Hamiltonian\newline
$\hat{H}_\mathrm{aa}=\int\hat{\Psi}^\dag(\mathbf{r})\hat{\Psi}^\dag(\mathbf{r'})\mathcal{V}(\mathbf{r},\mathbf{r'})\hat{\Psi}(\mathbf{r'})\hat{\Psi}(\mathbf{r})\mathrm{d}^3\mathbf{r}\mathrm{d}^3\mathbf{r'}$
with atomic field operator $\hat{\Psi}(\mathbf{r})$. This
Hamiltonian is obtained from the dispersive atom-light interaction
Hamiltonian by adiabatically eliminating the fast cavity field
dynamics \cite{SOM,maschler2005}. The interaction potential has
the form
\begin{equation}\label{atom-atom-int}
\mathcal{V}(\mathbf{r}, \mathbf{r}')= V \cos(kx)\cos(kz) \cos(kx')\cos(kz')
\end{equation}
and describes the ac Stark shift experienced by an atom at
position $\mathbf{r}$ as a result of the cavity field induced by a
second atom at position $\mathbf{r}'$. As a consequence of the
interference between the mediating cavity field and the transverse
pump field, the spatial dependence of the interaction is
determined by the corresponding mode functions $\cos(kx)$ and
$\cos(kz)$, where $k=2\pi/\lambda$ denotes the optical wavevector
(Fig.~1A). The interaction strength $V =\hbar\eta^2 /
\tilde{\Delta}_\mathrm{c}$ depends on the two-photon Rabi
frequency $\eta$ which can be tuned experimentally via the
transverse pump power $P$ \cite{SOM}. The sign of $V$ is
determined by the detuning
$\tilde{\Delta}_\mathrm{c}=\omega_\mathrm{p}-\tilde{\omega}_\mathrm{c}$
between the transverse pump laser frequency $\omega_\mathrm{p}$
and the dispersively shifted cavity resonance
$\tilde{\omega}_\mathrm{c}$. In the experiment
$|\tilde{\Delta}_\mathrm{c}|$ is large compared to the cavity
linewidth $2\kappa$.

For negative $V$, the cavity-mediated interaction induces density
correlations in the atomic cloud with spatial periodicity of
$\lambda$ along the pump and cavity direction. In momentum space,
this corresponds to the creation and annihilation of pairs of
correlated atoms in the momentum mode $|\mathrm{e}\rangle$, which
is the symmetric superposition of the four momentum states $|p_x,
p_z\rangle = |\pm\hbar k, \pm \hbar k\rangle$. For a
macroscopically populated zero-momentum mode containing $N$ atoms,
this is described by the Hamiltonian \cite{SOM}
\begin{equation}\label{Ising}
\hat{H} =  2 E_\mathrm{r}\hat{c}^{\dag}\hat{c} +
\frac{NV}{4}(\hat{c}^{\dag}+\hat{c})^{2},
\end{equation}
where $\hat{c}^{\dag}$ creates particles in mode
$|\mathrm{e}\rangle$. The first term in Eq.~\ref{Ising}
corresponds to the kinetic energy of atoms in mode
$|\mathrm{e}\rangle$, with the recoil energy $E_\mathrm{r} =
\frac{\hbar^2k^2}{2m}$ and the atomic mass $m$.

%%%%%%%%%%%%
\begin{figure}
\centering
\includegraphics[width=\columnwidth]{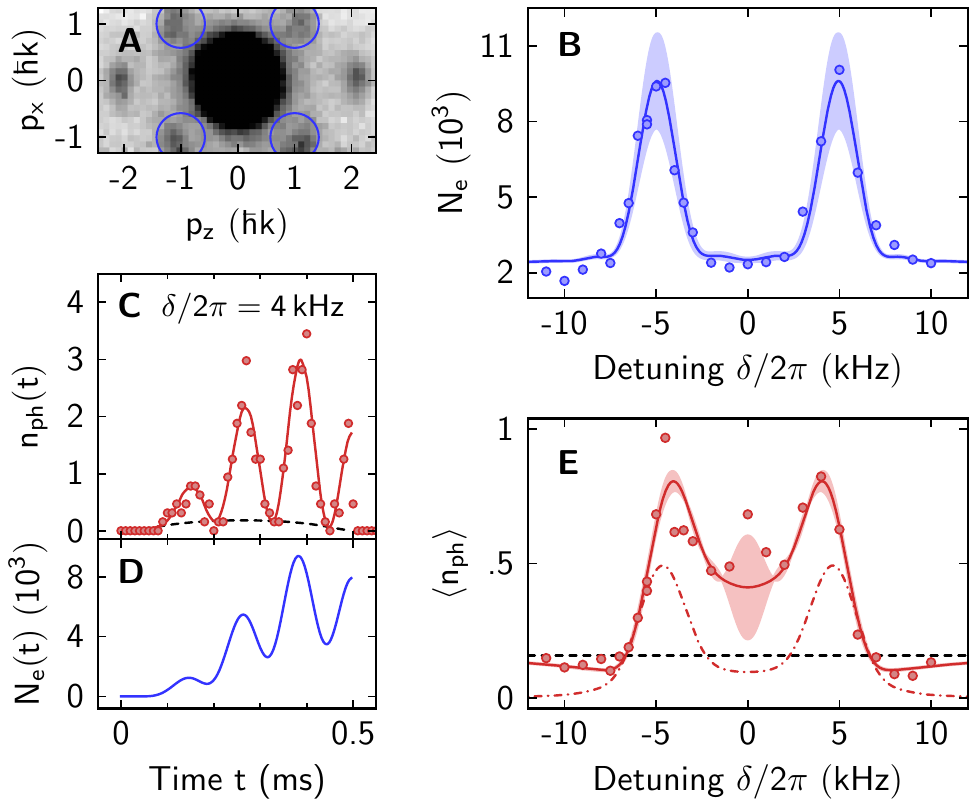}{}
\caption{
Probing the excitation spectrum in the normal phase
($P=0.65P_{\mathrm{cr}})$. (\textbf{A}) Absorption image of the
atomic cloud after probing the cavity for $\unit[0.5]{ms}$ and
subsequent ballistic expansion over $\unit[7]{ms}$. (\textbf{B})
Excited atomic population $N_\mathrm{e}$ with momenta $(\pm \hbar
k, \pm \hbar k)$, deduced from the regions enclosed by blue
circles in A, as a function of the probe-pump detuning $\delta$
(circles). The solid line shows a fit, based on the theoretical
model (see text). The background in $N_\mathrm{e}$ originates from
a residual thermal component extending into the detection region.
(\textbf{C}) Recorded (circles) and fitted (solid line) mean
intracavity photon number $n_\mathrm{ph}$ during probing. The
dashed line indicates the intracavity photon number of the probe
pulse, obtained in the absence of atoms. (\textbf{D}) Temporal
evolution of $N_\mathrm{e}(t)$ inferred theoretically from the fit
shown in C. (\textbf{E}) Mean intracavity photon number $\langle
n_\mathrm{ph}\rangle$, averaged over the probe interval, as a
function of $\delta$ (circles). The fit (solid line) takes into
account the calibrated photon number of the probe pulse (dashed
line). From the recorded photon signal we deduce the photon number
originating from Bragg scattering off the created excitations
(dashed-dotted line). The shading in B and E indicates the
fluctuations caused by variations of the relative phase $\varphi$
between different experimental runs. The detuning of the pump beam
from the empty cavity resonance was $\Delta_{c} = \unit[-2\pi
\times 18]{MHz}$ and the total atom number $N=1.6(2)\times 10^5$.
The damping constant $\gamma$ was set to $\unit[0.6]{kHz}$ (see
text).} \label{figure2}
\end{figure}
%%%%%%%%%%%%%

The elementary excitations of the Hamiltonian $\hat{H}$ are of
collective nature and their eigenenergy $E_\mathrm{s}$ softens for
increasing interaction strength $V$ (Fig.~1B). The excitation
energy $E_\mathrm{s}$ vanishes at a critical interaction strength
$V_\mathrm{cr}$, where interaction and kinetic energy are
balanced, i.e.~$N|V_\mathrm{cr}| = 2 E_\mathrm{r}$. This marks the
transition point between a normal and a supersolid phase. In the
supersolid phase, a macroscopic occupation of the mode
$|\mathrm{e}\rangle$ gives rise to a checkerboard pattern in the
atomic density distribution, which is accompanied by the build-up
of a coherent cavity field amplitude $\alpha_0$. The energy of
collective excitations rises again with increasing interaction
strength and approaches the single-particle excitation energy of
the induced optical checkerboard potential.

To measure the excitation spectrum of the system at momenta $(\pm
\hbar k, \pm \hbar k) $, we perform a variant of Bragg
spectroscopy \cite{stenger1999,steinhauer2002}. After preparing
the system at a given interaction strength, the cavity field is
excited with a weak probe pulse along the $x$-axis (Fig.~1A).
Interference between the cavity probe field and the transverse
pump field results in an amplitude-modulated lattice potential
$\eta \sqrt{n_\mathrm{pr}} \cos(\delta
t+\varphi)\cos(kx)\cos(kz)$, with probe-pump detuning
$\delta=\omega_{\mathrm{pr}}-\omega_{\mathrm{p}}$, relative phase
$\varphi$ and mean intracavity photon number $n_\mathrm{pr}$ of
the probe field. In momentum space, this corresponds to the
perturbation $\hat{H}_\mathrm{pr}= \eta \sqrt{n_\mathrm{pr}
N}\cos(\delta t + \varphi) (\hat{c}^{\dag}+\hat{c})$. After
probing, all laser fields are switched off. This projects the
created collective excitations onto the free-space momentum states
$\lvert \pm \hbar k, \pm \hbar k\rangle $, which are detected via
absorption imaging subsequent to ballistic expansion (Fig.~2A). A
typical resonance curve of the excited momentum state population
$N_\mathrm{e} = \langle \hat{c}^{\dag}\hat{c}\rangle$ as a
function of $\delta$ is shown in Fig.~2B. Clear resonances are
revealed both for positive and negative probe-pump detuning,
corresponding to stimulated scattering of probe photons into the
pump field and vice versa. The corresponding resonance frequency
$E_\mathrm{s}/\hbar$ is obtained from a fit (solid line) based on
the model description Eq.~\ref{Ising} \cite{SOM}.

The measured excitation spectrum as a function of pump power $P$
is displayed in Fig.~3. When increasing the interaction strength
in the normal phase towards the critical point, the excitation
energy $E_\mathrm{s}$ exhibits a distinct softening. In contrast,
for positive $V$, the excitation gap is observed to increase with
interaction strength in accordance with the absence of a phase
transition. The increasing gap reflects the tendency of the
cavity-mediated interactions to suppress density fluctuations with
$\lambda$-periodicity.

We monitor the cavity output light on a single-photon counting
module. This gives us real-time access to the formation of
collective excitations during probing. Because of matter-wave
interference between the ground-state component of the condensate
and the created excitations, a spatial density modulation evolves
in time. Transverse pump light is Bragg scattered by this density
modulation into the cavity mode. The output from the cavity
therefore follows the oscillatory evolution of this density
modulation (Fig. 2C). The total intracavity photon number,
averaged over the probe interval, again exhibits a double
resonance (Fig.~2E), whose characteristic shape has its origin in
interference between the probe field (dashed) and the Bragg
scattered pump field (dashed-dotted). This provides a second
method to determine the resonance frequency $E_\mathrm{s}/\hbar$,
as shown in Fig.~3.

We model the dynamics of the system during probing by evolving the
Hamiltonian $\hat{H}+\hat{H}_\mathrm{pr}$ in time, taking into
account the recorded shape of the probe pulse, see Fig.~2C.
Quantitative agreement with the data is obtained by including a
phenomenological damping rate $\gamma$ in the time evolution of
$\hat{c}$ \cite{SOM}. The corresponding time evolution of the
momentum population $N_\mathrm{e}(t)$ is shown in Fig.~2D.
Depending on the relative phase $\varphi$, which is not controlled
in the experiment, the phase of the excited density oscillation
varies between different experimental runs. This leads to an
intrinsic fluctuation of the quantities $N_\mathrm{e}$ and
$\langle n_\mathrm{ph}\rangle$, as indicated by the shaded areas
in Fig.~2B and E.

In the supersolid phase, the perturbation creates excitations on
top of the macroscopic steady-state population in momentum mode
$|\mathrm{e}\rangle$ \cite{baumann2010}. Because of matter-wave
interference, the detected population $N_\mathrm{e}$ is strongly
affected by fluctuations of the relative phase $\varphi$. Yet, the
variance of $N_\mathrm{e}$ around the steady-state value, deduced
from several experimental runs, shows clear resonances.
Alternatively, we use the oscillation amplitude of the intracavity
photon number around the steady-state value $|\alpha_0|^2$ to
extract the excitation energy $E_\mathrm{s}$ \cite{SOM}. As shown
in Fig.~3, the excitation energies deduced from these two signals
rise again with pump power $P$ and provide a consistent picture.

%%%%%%%%%%%%
\begin{figure}
\centering
\includegraphics[width=\columnwidth]{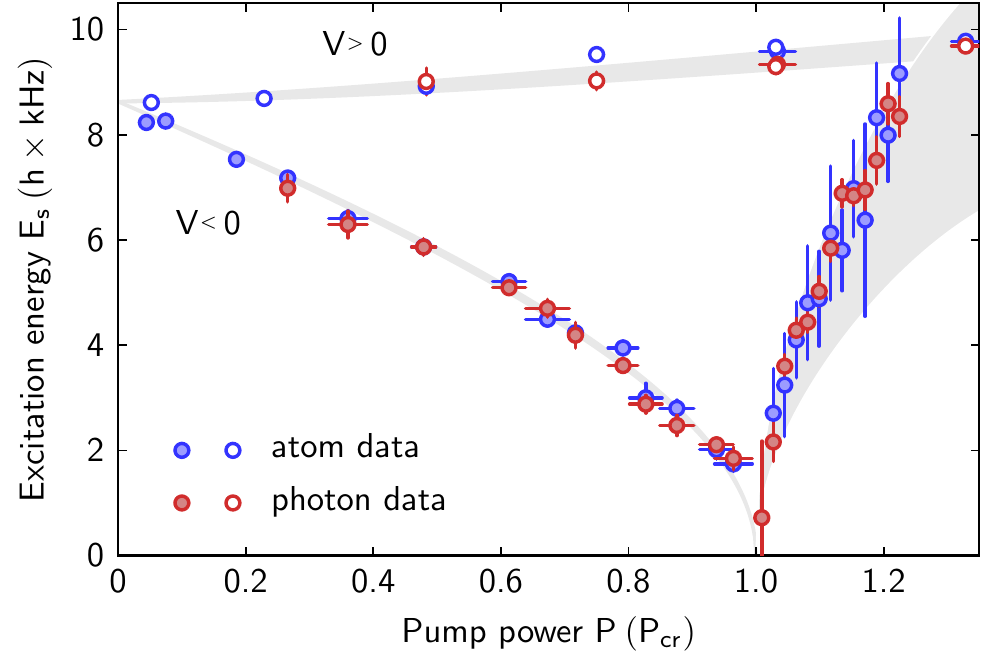}{}
\caption{
Excitation spectrum. Measured resonance frequencies
$E_\mathrm{s}/h$, obtained from atomic ($N_\mathrm{e}$) and
photonic ($\langle n_\mathrm{ph}\rangle$) signals, are shown in
blue and red, respectively, for positive (open circles) and
negative (filled circles) interaction strength $V$. Gray shading
shows the theoretical prediction including experimental
uncertainties \cite{SOM}. The experimental parameters are
$N=1.7(2)\times10^5$, $\Delta_\mathrm{c}=-2\pi \times (19.8, 23.2)
\mathrm{MHz}$ for $V<0$ (normal and supersolid phase) and
$\Delta_\mathrm{c}=+2\pi \times 15.1\,\mathrm{MHz}$ for $V>0$. For
$V>0$, the pump power is scaled in terms of the critical pump
power observed for $\Delta_\mathrm{c}=-2\pi \times
19.7\,\mathrm{MHz}$.
}\label{figure3}
\end{figure}
%%%%%%%%%%%%%

The observed excitation spectrum (Fig.~3) is in qualitative
agreement with a soft mode energy $E_{\mathrm{s}} = 2E_\mathrm{r}
\sqrt{1+V/|V_\mathrm{cr}|}$ which follows from Eq.~\ref{Ising}.
Quantitative agreement is obtained by further taking into account
the lattice potential of the transverse pump beam as well as
contact interactions between colliding atoms \cite{SOM}.
Accordingly, the single-particle mode $|\mathrm{e}\rangle$ in
Eq.~\ref{Ising} is replaced by a Bogoliubov mode $|1\rangle$ and
the interaction energy gets renormalized by the matrix element of
$\mathcal{V}(\mathbf{r},\mathbf{r}')$ between the condensate mode
and the excited state $\lvert 1 \rangle$. In the normal phase, the
state $\lvert 1 \rangle$ lies in the lowest energy band of the
transverse lattice potential with quasi-momentum $(\pm \hbar k,
\pm \hbar k)$ and bare energy $E_{1}$. For $P = 0$ the excitation
energy $E_\mathrm{s}$ reaches the Bogoliubov energy
$E_\mathrm{1}=h\times\unit[8.7]{kHz}$, which is shifted by the
mean-field energy with respect to the kinetic energy $2
E_\mathrm{r} = h\times\unit[7.5]{kHz}$. In the supersolid phase,
the Bogoliubov mode $|1\rangle$, which is dominantly affected by
the long-range interactions, lies in a higher energy band at zero
quasi-momentum of the emerging checkerboard lattice potential. As
a result of competition between the increasing band energy and the
renormalized interaction energy, the excitation energy
$E_\mathrm{s}$ rises again.

%%%%%%%%%%%%
\begin{figure}
\centering
\includegraphics[width=\columnwidth]{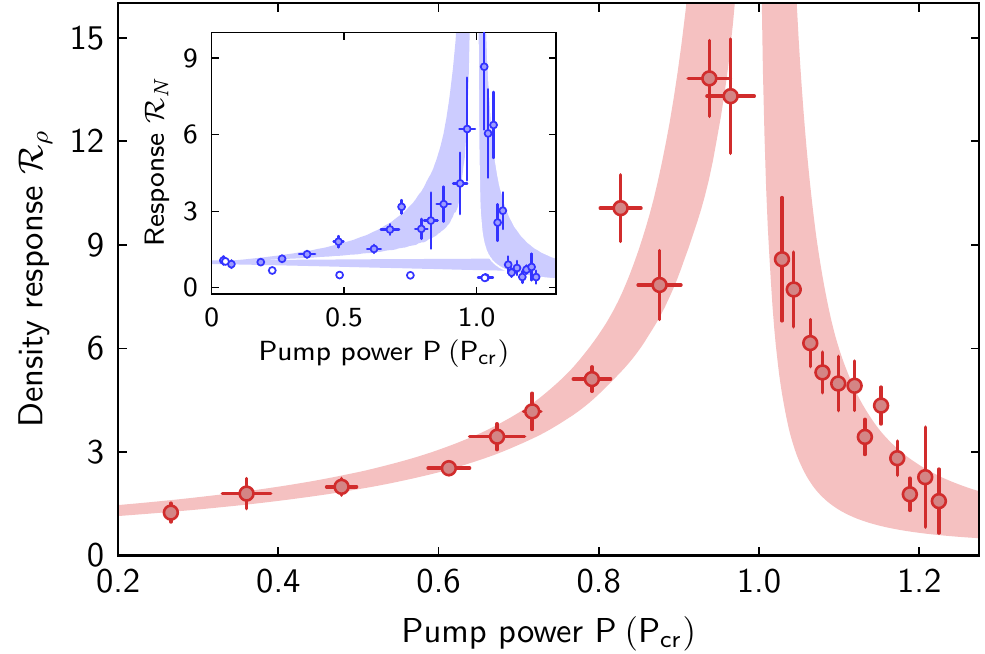}{}
\caption{
Response to density perturbations. Shown is the density response
in the normal and supersolid phase, normalized to the
non-interacting case ($P=0$). The shaded areas show the
theoretical prediction including experimental uncertainties and
fluctuations caused by the uncontrolled relative phase $\varphi$.
The atomic damping constant $\gamma =
2\pi\times\unit[0.5(1)]{kHz}$ was obtained from the fit to the
data in the normal phase. The inset shows the response of the
excited momentum state population $N_\mathrm{e}$ for negative
(filled circles) and positive (open circles) interaction strength
$V$. The experimental parameters were the same as in Fig.~3.
}\label{figure4}
\end{figure}
%%%%%%%%%%%%%

Another signature for a softened dispersion relation at finite
momentum is provided by the susceptibility to external density
perturbations. In terms of the static structure factor $S(k)$,
which measures the density-density response, this is expressed by
Feynman's relation $S(k) = (\hbar k^2/2m)/\omega(k)$, where
$\omega(k)$ denotes the dispersion of a homogeneous system
\cite{nozieres1990}. To determine the density response in our
system, we quantify the amount of pump light which is Bragg
scattered into the cavity. This allows us to measure the
probe-induced density modulation $\langle
\delta\hat{\rho}_\mathrm{e}\rangle$, where $\delta
\hat{\rho}_\mathrm{e}$ denotes the density fluctuation operator at
momenta $(\pm \hbar k, \pm \hbar k)$. We define the corresponding
quadratic density response function $\mathcal{R}_\rho$ as the
spectral weight of $\langle \delta\hat{\rho}_\mathrm{e}\rangle^2$,
normalized to the pump power $P$ and the probe pulse area $\langle
n_\mathrm{pr}\rangle$. The response function $\mathcal{R}_N$ of
the atomic population $N_\mathrm{e}$ in the excited momentum state
$|\mathrm{e}\rangle$ is defined in a similar way \cite{SOM}.

Approaching the critical point from below, a strongly increasing
density response is observed (Fig.~4). This indicates the presence
of enhanced density fluctuations with $\lambda$-periodic
correlations in the unperturbed system. The data is in good
agreement with our model (shading), which predicts
$\mathcal{R}_{\rho}$ to scale as $(E_1/E_\mathrm{s})^2$ in the
normal phase, in accordance with Feynman's relation. In the
supersolid phase, the density response decreases again as a
function of $P$, as expected for the increasing excitation energy
(Fig.~3). A similar behavior is observed for the response in the
momentum state population $N_\mathrm{e}$ (Fig.~4, inset), which
exhibits a larger sensitivity to variations of the relative phase
$\varphi$ \cite{SOM}. For positive interaction $V$ (Fig.~4,
inset), the response of the system is reduced with respect to the
non-interacting case, $V=0$.

We have observed a roton-type mode softening causing a
superfluid-to-supersolid transition in a model system for
long-range interactions. Increasingly complex spatial structures
of long-range atom-atom interactions
\cite{gopalakrishnan2009,strack2011,gopalakrishnan2011} can be
tailored by extending the experimental setup to multiple cavity
modes.

% Your references go at the end of the main text, and before the
% figures.  For this document we've used BibTeX, the .bib file
% scibib.bib, and the .bst file Science.bst.  The package scicite.sty
% was included to format the reference numbers according to *Science*
% style.

\section*{Acknowledgment} We thank H. Tureci, P. Domokos and H. Ritsch for
stimulating discussions. Financial funding from SQMS (ERC advanced
grant), NAME-QUAM (EU, FET open), NCCR-QSIT and ESF (POLATOM) is
acknowledged.

%\bibliography{ours}
%\bibliographystyle{Science}
%\bibliographystyle{apsrev4-1}

%

\pagebreak

\begin{appendix}
\section*{Supplementary Materials}

We give details of the experimental setup and provide a
theoretical description of the dispersively coupled
condensate-cavity system. An effective Hamiltonian for the
cavity-mediated long-range interaction is derived. Based on a
mean-field description, we numerically calculate the steady state
of the system including contact interactions and the transverse
lattice potential. We deduce in a Bogoliubov approach the energy
of collective excitations, which softens towards the critical
point. The time evolution of the system during probing is derived
and a diverging response to density perturbations at the phase
transition is found.

\subsection*{Experimental details}

The atoms are prepared in the hyperfine state $(F, m_F) = (1, -1)$
with respect to a quantization axis pointing along the cavity
axis, where $F$ is the total angular momentum and $m_F$ the
magnetic quantum number. The transverse pump laser with wavelength
$\lambda = \unit[784.5]{nm}$ is linearly polarized along the
$y$-axis (see Fig.~1 in the main text) and off-resonantly drives,
via the atoms, two degenerate cavity TEM$_{00}$ modes with
circular polarizations $\epsilon_+$ and $\epsilon_-$. The ratio of
the corresponding two-photon Rabi frequencies is given by
$\eta_{+}/\eta_{-} = 3.26/1.25$, where all allowed dipole
transitions in the $D_1$ and $D_2$ lines of $^{87}$Rb have been
taken into account. The maximum dispersive shift of the two cavity
modes induced by a single maximally coupled atom is $U_0^{+} =
2\pi\times \unit[87]{Hz}$ and $U_0^{-} = 2\pi\times
\unit[33]{Hz}$, respectively. The transverse pump laser induces an
optical lattice potential along the $z$-axis with periodicity of
$\lambda/2$. Its depth $V_\mathrm{p}$ is calibrated using
Raman-Nath diffraction \cite{morsch2006} and takes for our
experimental parameters a value of $3 E_\mathrm{r}$ at the
critical point, with recoil energy $E_\mathrm{r} = \frac{\hbar^2
k^2}{2m}$, wavevector $k = 2\pi/\lambda$, and atomic mass $m$. In
the theoretical analysis, the Gaussian envelopes of the pump and
cavity fields along the transverse directions \cite{baumann2010}
are effectively taken into account by weighted averages of
$V_\mathrm{p}$, $\eta_\pm$ and $U_0^\pm$ over the spatial extent
of the atomic cloud. The Thomas-Fermi radii of the condensate in
the external harmonic trapping potential are given by $(R_x, R_y,
R_z) = (3.5,18.3,3.7)$ \textmu m, assuming an atom number of $N =
1.65 \times 10^5$.

The length ($\unit[176]{\mu m}$) of the cavity is actively
stabilized using a laser with a wavelength of $\unit[830]{nm}$,
which is referenced onto the transverse pump laser. The depth of
the resulting intracavity lattice potential was measured to be
$0.04(2) E_\mathrm{r}$, and is neglected in the theoretical
analysis. Transverse pump light and cavity probe light propagate
through independent optical fibers, resulting in a variation of
their relative phase $\varphi$ between different experimental
runs. The cavity output light is monitored on a single-photon
counting module with an overall detection efficiency of
intracavity photons of $4(1)\%$. For the data taken in the normal
phase, the critical pump power $P_\mathrm{cr}$ was deduced from
the intracavity photon number monitored during independent sweeps
across the phase transition.

\subsection*{Theoretical description of the coupled condensate-cavity system}
\emph{The coupled condensate-cavity system is described in a
many-body formalism following
Ref.~\cite{maschler2005,maschler2008}. By adiabatically
eliminating the fast cavity field dynamics, we derive an effective
Hamiltonian, which describes the long-range atom-atom
interaction.}

After adiabatically eliminating the electronically excited states,
the transversally driven condensate-cavity system is described by
the many-body Hamiltonian
$\hat{H}=\hat{H}_\mathrm{c}+\hat{H}_\mathrm{a}+\hat{H}_\mathrm{a-c}$,
where
\begin{equation}\label{eq:Hamiltonian-full}\tag{S1}
\begin{split}
\hat{H}_\mathrm{c}=&-\hbar \Delta_\mathrm{c} \hat{a}^\dag \hat{a} \\
\hat{H}_\mathrm{a}=&\ \int d^3r \hat{\Psi}^\dag(\mathbf{r})\bigg[\frac{\mathbf{p}^2}{2m}  + V_\mathrm{p}\cos^2(kz)\\
 &+ \frac{g}{2}\,\hat{\Psi}^\dag(\mathbf{r})\hat{\Psi}(\mathbf{r}) \bigg] \hat{\Psi}(\mathbf{r})  \\
\hat{H}_\mathrm{a-c}=& \int d^3r \hat{\Psi}^\dag(\mathbf{r}) \big[
\hbar \eta \cos(kx) \cos(kz) (\hat{a}+\hat{a}^\dag)  \\
&+\hbar U_0 \cos^2(kx) \hat{a}^\dag \hat{a} \big]  \hat{\Psi}(\mathbf{r})\,, 
\end{split}
\end{equation}
with bosonic atomic field operator $\hat{\Psi}(\mathbf{r})$ and
photon operators $\hat{a}$ and $\hat{a}^\dag$. To keep the notation
simple we describe only one of the two circularly polarized
TEM$_{00}$ cavity modes. Final results will be given for the case of
two cavity modes.

In equation \eqref{eq:Hamiltonian-full}, $\hat{H}_\mathrm{c}$
describes the dynamics of a single TEM$_{00}$ cavity mode with
spatial mode function $\cos(kx)$, whose frequency
$\omega_\mathrm{c}$ is detuned by $\Delta_\mathrm{c} =
\omega_\mathrm{p}- \omega_\mathrm{c}$ from the pump laser
frequency $\omega_\mathrm{p}$. The term $\hat{H}_\mathrm{a}$
captures the atomic evolution in the transverse optical lattice
potential with depth $V_\mathrm{p}$, including contact
interactions with strength $g= \frac{4\pi \hbar^2a}{m}$, where $a$
denotes the s-wave scattering length. The interaction between the
atoms and the pump and cavity light fields is governed by
$\hat{H}_\mathrm{a-c}$. The first term describes light scattering
between pump and cavity field at a rate which is determined by the
maximum two-photon Rabi frequency $\eta$. The second term accounts
for the dispersive shift of the cavity resonance frequency with
light-shift $U_0$ of a single maximally coupled atom.

As the cavity field reaches a steady-state on a time scale fast
compared to atomic motion, its equation of motion can be formally
solved, yielding
\begin{equation}\label{eq:steady-state-cavity-field}\tag{S2}
\hat{a} = \frac{\eta
\hat{\Theta}}{(\Delta_\mathrm{c}-U_0\hat{\mathcal{B}}) + i\kappa}
\end{equation}
with the cavity field decay rate $\kappa=2\pi\times
1.25~\mathrm{MHz}$. Due to Bragg scattering of pump light, the
intracavity field amplitude is proportional to the order parameter
$\hat{\Theta} = \int \mathrm{d}^3r \,\hat{\Psi}^\dag (\mathbf{r})
\cos(kx) \cos(kz) \hat{\Psi}(\mathbf{r})$ which measures the
atomic density modulation on the checkerboard pattern
$\cos(kx)\cos(kz)$. The overall dispersive shift of the empty
cavity resonance caused by the presence of the atoms is
proportional to the bunching parameter $\hat{\mathcal{B}} = \int
\mathrm{d}^3r\,
\hat{\Psi}^\dag(\mathbf{r})\cos^2(kx)\hat{\Psi}(\mathbf{r})$.

After eliminating the steady-state cavity field of
Eq.~\eqref{eq:steady-state-cavity-field} from Hamiltonian
Eq.~\eqref{eq:Hamiltonian-full}, an effective Hamiltonian
description is obtained (see main text, Eq.~1):
\begin{displaymath}\label{eq:interaction-hamiltonian}\tag{S3}
\begin{array}{l r}
\hat{H}_\mathrm{eff}=& \hat{H}_\mathrm{a} +   V\int \mathrm{d}^3\mathbf{r}\mathrm{d}^3\mathbf{r'} \hat{\Psi}^\dag(\mathbf{r})\hat{\Psi}^\dag(\mathbf{r'}) \cos(kx)\cos(kz) \\
\\[-5pt]
& \times\cos(kx')\cos(kz')
\hat{\Psi}(\mathbf{r}) \hat{\Psi}(\mathbf{r'}).
\end{array}
\end{displaymath}
The strength $V$ of this cavity-meditated atom-atom interaction is
given by
$V=\hbar\frac{\eta^2\tilde{\Delta}_\mathrm{c}}{\tilde{\Delta}_\mathrm{c}^2+\kappa^2}\approx\hbar\frac{\eta^2}{\tilde{\Delta}_\mathrm{c}}$,
where the detuning of the pump laser from the dispersively shifted
cavity resonance $\tilde{\Delta}_\mathrm{c} = \Delta_\mathrm{c} -
U_0\mathcal{B}_0$ was taken to be large compared to the cavity
half-linewidth $\kappa$. Here, $\mathcal{B}_0 =
\langle\hat{\mathcal{B}}\rangle$ denotes the bunching parameter in
the steady state. In a quantized picture, $N V/\hbar$ corresponds
to the rate at which cavity photons are exchanged between atoms,
as shown exemplarily in the zoom of Fig.~1A in the main text.

The effective Hamiltonian Eq.~\eqref{eq:interaction-hamiltonian}
describes a closed system and neglects dynamical and quantum
backaction effects originating from cavity decay and cavity input
noise. This is justified on short timescales as long as
$|\tilde{\Delta}_{\mathrm{c}}| \gg \kappa \gg E_{\mathrm{r}} /
\hbar$ \cite{nagy2010}.

\subsection*{Mean-field description in the steady state}
\emph{Based on a mean-field description \cite{nagy2008}, we derive a
numerical solution of the steady state of the system.}

A mean-field description of the system is obtained by formally
replacing the operators $\hat{\Psi}$ and $\hat{a}$ in
Eq.~\eqref{eq:Hamiltonian-full} with the atomic mean-field
$\sqrt{N}\psi_{0}$ and the coherent cavity amplitude $\alpha_{0}$,
respectively. Their steady-state values are determined by the
non-local Gross-Pitaevskii equation
\begin{equation}\label{eq:GPE}\tag{S4}
\begin{split}
\mu_0 \psi_0(x,z) &= \bigg(
\frac{-\hbar^2}{2m}(\partial^2_x+\partial^2_z) +V_\mathrm{p}(z)+ \hbar U_0(x)|\alpha_0|^2 \\
& + \hbar\eta(x,z)(\alpha_0+\alpha_0^*)  +
g_\mathrm{2D}|\psi_0|^2 \bigg)\psi_0(x,z)
\end{split}
\end{equation}
with
$\alpha_0=\frac{\eta\Theta_0}{\tilde{\Delta}_\mathrm{c}+i\kappa}$
and the chemical potential $\mu_{0}$. Here, we introduced the
steady state order parameter $\Theta_0=N \langle \psi_0
|\cos(kx)\cos(kz)| \psi_0 \rangle$ and bunching parameter
$\mathcal{B}_0=N \langle \psi_0 |\cos^2(kx)|\psi_0 \rangle$, as
well as the notations $V_\mathrm{p}(z) = V_\mathrm{p} \cos^2(kz)$,
$\eta(x,z) = \eta \cos(kx)\cos(kz)$ and $U_{0}(x) =
U_{0}\cos^2(kx)$. We reduced the description in Eq.~\eqref{eq:GPE}
to the pump and cavity directions, assuming a homogeneous
condensate density along the weakly confined $y$-axis. The contact
interaction strength is accordingly replaced by $g_\mathrm{2D} =
\lambda^2\bar{n} g$ with average 3D condensate density $\bar{n}$
and the normalization condition $\int_{0}^\lambda\mathrm{d}x
\int_{0}^\lambda \mathrm{d}z \,|\psi_{0}|^2 =1$
\cite{kraemer2003}.

%For negative $V$, the long-range interaction term in
%Eq.~\eqref{eq:interaction-hamiltonian} drives a second order phase
%transition from a normal to a supersolid phase with
%$\lambda$-periodic density modulation accompanied by off-diagonal
%long-range order, see ref. \cite{baumann2010}.

For negative $\tilde{\Delta}_\mathrm{c}$, Eq.~\eqref{eq:GPE}
exhibits a dynamical instability above a critical transverse pump
power $P_\mathrm{cr}$, driving the system from a normal phase into a
supersolid phase with $\lambda$-periodic density modulation. In the
normal phase, $P<P_\mathrm{cr}$, the condensate density is flat
along the cavity axis, resulting in a vanishing order parameter,
$\Theta_0=0$, and cavity field amplitude, $\alpha_0=0$. The
mean-field solution $\psi_0$ is given by the lowest energy Bloch
state in the shallow optical lattice potential of the transverse
pump field. In the supersolid phase, $P>P_\mathrm{cr}$, the atomic
cloud exhibits a $\lambda$-periodic density modulation both along
the transverse and the cavity direction. Correspondingly, the order
parameter takes a finite value, $\Theta_0\neq0$, and light
scattering off the diagonal Bragg planes results in a coherent
cavity field amplitude, $\alpha_0\neq0$. The mean-field solution
$\psi_0$ is given by the minimal energy state in the two-dimensional
lattice potential originating from interference between the
transverse and cavity field. In the supersolid phase, the emergent
checkerboard density modulation locks to one of two possible
sublattices, which are spatially shifted by $\lambda/2$ and have
opposite signs of $\Theta_0$ and $\alpha_0$. As both solutions share
the same excitation spectrum, we assume $\Theta_0
> 0$ in the following.

We numerically find the ground state $\psi_0$ of the system by
propagating Eq.~\eqref{eq:GPE} in imaginary time, using a
computational cell of size $\lambda^2$ with periodic boundary
conditions. We include the recorded steady state mean intracavity
photon number $|\alpha_0|^2$ and the experimentally calibrated
lattice depth $V_\mathrm{p}$ in the calculations.

\subsection*{Deriving the collective excitation spectrum}
\emph{We describe collective excitations on top of the mean-field
solution. To this end, we calculate the Bogoliubov  excitations in
the steady-state lattice potential and identify a single
excitation mode which is dominantly affected by the long-range
interactions. We diagonalize the resulting truncated Hamiltonian
and deduce the energy of collective excitations.}

Following Ref.~\cite{nagy2008}, we expand the atomic and cavity
field operators around their mean-field solution $(\psi_0,
\alpha_{0})$ according to
\begin{align}\label{eq:steady-state-expansion}
&\hat{\Psi}=(\sqrt{N}\psi_0+\delta\hat{\Psi})e^{-it\mu_0/\hbar}\notag\\
&\hat{a}=\alpha_0+\delta\hat{a}.\tag{S5}
\end{align}
Here, the linearized cavity field fluctuation operator
\begin{equation}\label{eq:delta-a}\tag{S6}
\delta\hat{a} = \frac{\eta}{\tilde{\Delta}_\mathrm{c}+i\kappa}
\left[ \delta\hat{\Theta} +
\frac{\Theta_{0}U_0}{\tilde{\Delta}_\mathrm{c}+i\kappa}\delta\hat{\mathcal{B}}
\right ].
\end{equation}
is given in terms of the operators $\delta\hat{\Theta} =\sqrt{N}
\int \mathrm{d}^3r\,
(\delta\hat{\Psi}^{\dag}+\delta\hat{\Psi})\cos(kx)\cos(kz)\psi_0$
and $\delta\hat{\mathcal{B}}=\sqrt{N}\int \mathrm{d}^3r\,
(\delta\hat{\Psi}^{\dag}+\delta\hat{\Psi})\cos^2(kx)\psi_0$,
describing fluctuations of the order and bunching parameter around
their steady-state values $\Theta_{0}$ and $\mathcal{B}_{0}$,
taking $\psi_0$ to be real-valued.

After eliminating the cavity field fluctuation $\delta\hat{a}$,
assuming $\tilde{\Delta}_\mathrm{c}^2\gg\kappa^2$, and keeping
only linear terms in the equation of motion for
$\delta\hat{\Psi}$, we arrive at a quadratic Hamiltonian
\begin{equation}\label{eq:eff-ham-excs}\tag{S7}
\hat{H}_\mathrm{exc}= \hat{H}_0 + V \biggl[ \delta\hat{\Theta} +
\frac{\Theta_0U_0
}{\tilde{\Delta}_\mathrm{c}}\delta\hat{\mathcal{B}} \biggl]^2
\end{equation}
describing the dynamics of atomic fluctuations around the mean-field
solution. The first term $\hat{H}_0$ captures the dynamics in the
static steady-state lattice potential including contact
interactions, and is given by
\begin{equation}\label{eq:ham-with-alpha0}\tag{S8}
\begin{split}
\hat{H}_0 = & \int d^3r \,\,\delta\hat{\Psi}^\dag \bigg[ \frac{-\hbar^2}{2m}(\partial^2_x+\partial^2_z) +V_\mathrm{p}(z) \\
& +\hbar\eta(x,z) (\alpha_0+\alpha_0^*) + \hbar U_0(x) |\alpha_0|^2 \bigg] \delta\hat{\Psi} \\
&+ \frac{1}{2}\,g_\mathrm{2D}\,\psi_0^2\left(\delta\hat{\Psi}^2+(\delta\hat{\Psi}^\dag)^2\right)
+ 2g_\mathrm{2D}|\psi_0|^2\delta\hat{\Psi}^\dag\delta\hat{\Psi}.
\end{split}
\end{equation}
The  second term in Eq.~\eqref{eq:eff-ham-excs} describes how atomic
density fluctuations are affected by the cavity-mediated long-range
interaction.

In the normal phase, $\hat{H}_\mathrm{exc}$ is directly obtained
from Eq.~\eqref{eq:interaction-hamiltonian} by quadratic expansion
in $\delta\hat{\Psi}$ around the steady state. In the supersolid
phase, the dispersive cavity shift, which depends on the atomic
density distribution, gives rise to an additional long-range
interaction term with $\lambda/2$-periodicity along the cavity axis
\cite{maschler2008}, whose strength is proportional to the order
parameter $\Theta_0$. However, as the order parameter vanishes at
the critical point, solely the term $\delta\hat{\Theta}^2$ in
Eq.~\eqref{eq:eff-ham-excs} induces the normal to supersolid phase
transition.

To find the energy of collective excitations described by
$\hat{H}_\mathrm{exc}$, we first calculate the elementary
excitations of $\hat{H}_0$ using the Bogoliubov ansatz
\cite{pitaevskii2003,castin2001}
\begin{equation}\label{eq:bogoliubov}\tag{S9}
\delta\hat{\Psi}(\mathbf{r}) = \sum_{j} \left(
u_{j}(\mathbf{r})\hat{c}_{j} + v_{j}^*(\mathbf{r})\hat{c}_{j}^\dag
\right)\,,
\end{equation}
with band index $j$, Bogoliubov modes $u_{j}(\mathbf{r})$ and
$v_{j}(\mathbf{r})$, and corresponding mode operators $\hat{c}_{j}$,
fulfilling bosonic commutation relations. In this basis the
Hamiltonian $\hat{H}_\mathrm{exc}$ reads
\begin{equation}\label{eq:eff-ham-all-modes}\tag{S10}
\hat{H}_\mathrm{exc} = \sum_{j} \left(E_{j}
\hat{c}_{j}^{\dag}\hat{c}_{j} +
NV\chi_{j}(\hat{c}_{j}+\hat{c}^{\dag}_{j})^{2}\right),
\end{equation}
with Bogoliubov energies $E_{j}$ and interaction matrix elements
\begin{equation}\label{eq:chi-general}\tag{S11}
\chi_{j}=\Big\langle\psi_0 \Big |\cos(kx)\cos(kz)+
\frac{\Theta_0U_0}{\tilde{\Delta}_\mathrm{c}}\cos^2(kx) \Big\vert
u_{j} + v_{j} \Big\rangle^2.
\end{equation}
%%%%%%%%%%%%
\begin{figure}
\centering
\includegraphics[width=\columnwidth]{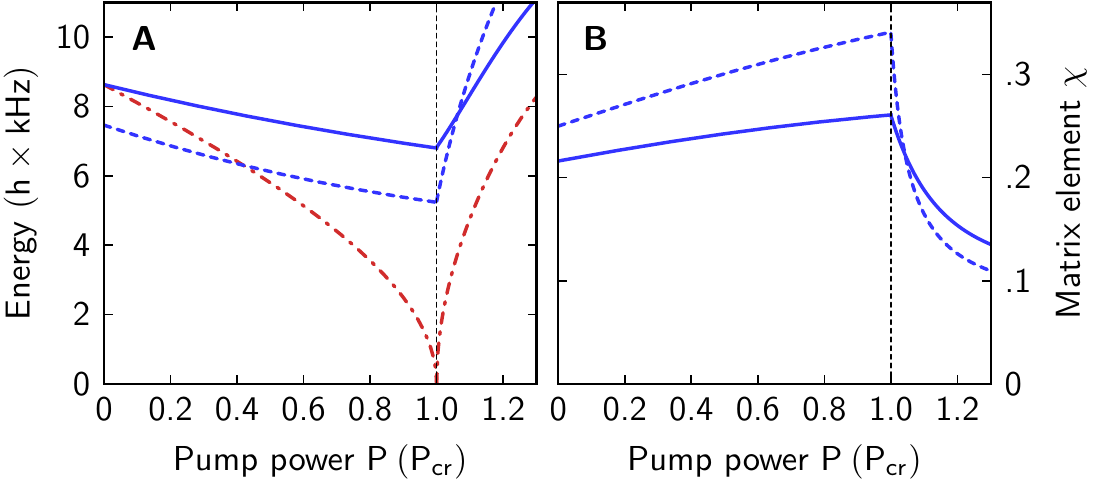}
\caption{(\textbf{A}) Bare energies (blue) of the excited state
with maximum matrix element $\chi$ in the normal and supersolid
phase. Dashed and solid blue lines correspond to the case of
vanishing and present contact atom-atom interactions. The red
dashed-dotted line shows the soft mode energy in the presence of
long-range and contact atom-atom interactions. (\textbf{B})
Maximum matrix element $\chi$ in the presence (solid) and absence
(dashed) of contact atom-atom interactions. }\label{figure1}
\end{figure}
%%%%%%%%%%%%%
From a numerical calculation of the matrix elements $\chi_{j}$ we
find in our parameter regime only a single Bogoliubov mode
exhibiting a relevant matrix element $\chi_{j}$ for energies up to
$15 E_{\textrm{r}}$. All other matrix elements $\chi_j$ are
suppressed by more than two orders of magnitude. In the normal
phase the maximally coupled excited mode lies in the lowest energy
band of the transverse lattice potential with quasi-momentum
$(q_x, q_z)=(\pm \hbar k, \pm \hbar k)$. In the supersolid phase
this state is folded into the third band at the center of the
Brillouin zone as a result of the emerging $\lambda$-periodicity.

In terms of the excited mode with dominant matrix element
$\chi_j$, which we denote in the following by the index $j=1$, the
effective Hamiltonian reduces to
\begin{equation}\label{eq:effective-hamiltonian}\tag{S12}
\hat{H}_\mathrm{exc} = E_\mathrm{1}
\hat{c}^{\dag}_\mathrm{1}\hat{c}_\mathrm{1} +
NV\chi(\hat{c}_\mathrm{1}+\hat{c}^{\dag}_\mathrm{1})^{2}
\end{equation}
with $\chi=\chi_1$. In the absence of the transverse lattice
potential and contact atom-atom interactions, $\hat{H}_\mathrm{exc}$
reduces in the normal phase to the Hamiltonian given in the main
text, Eq.~2.

We obtain the soft mode energy spectrum by diagonalizing Hamiltonian
$\hat{H}_\mathrm{exc}$ in terms of a second Bogoliubov
transformation $\hat{b} = \mu \hat{c}_1 + \nu \hat{c}_1^\dag$. Up to
a constant term, this yields
$\hat{H}_\mathrm{exc}=E_\mathrm{s}\hat{b}^\dag \hat{b}$ with soft
mode energy
\begin{equation}\label{eq:soft-mode}\tag{S13}
E_\mathrm{s}=E_\mathrm{1} \sqrt{1+\frac{4NV\chi}{E_1}}.
%\frac{V}{V_\mathrm{cr}}\frac{\chi}{\chi_\mathrm{cr}}\frac{E_\mathrm{1}^\mathrm{cr}}{E_\mathrm{1}}}.
\end{equation}

For $V<0$, $E_\mathrm{s}$ softens towards the phase transition and
vanishes at
$V_\mathrm{cr}=-E_\mathrm{1}^\mathrm{cr}/(4N\chi_\mathrm{cr})$ with
$(E_\mathrm{1}^\mathrm{cr},\chi_\mathrm{cr})=(E_1,\chi)|_{P=P_\mathrm{cr}}$.
In the supersolid phase, the excitation energy rises again due to
competition between the increasing Bogoliubov energy $E_1$ and the
decreasing interaction energy which is proportional to the matrix
element $\chi$ (see Fig. S\ref{figure1}).

Results for two circularly polarized cavity modes $\hat{a}_\pm$
are obtained by replacing $V \chi$ in
\eqref{eq:effective-hamiltonian} and \eqref{eq:soft-mode} by
$\sum_{n = \pm}V_n \chi_n$, with $V_\pm$ and $\chi_\pm$ deduced
accordingly from the cavity parameters $(\eta_\pm, U_0^\pm)$. The
calculated energy spectrum shown in Fig.~3 takes into account the
experimentally calibrated depth of the transverse lattice
potential and the measured steady-state intracavity photon number
$|\alpha_0|^2$ in the presence of atoms. Systematic uncertainties
of these quantities are estimated to be $10\%$ and $25\%$,
respectively (see shaded regions in Fig.~3).

\subsection*{Probing the collective excitation spectrum}
\emph{By integrating the equation of motion of the atomic field, we
predict the build-up of atomic excitations and of the cavity light
field during probing.}

To probe the excitation spectrum, we weakly drive the cavity field
with amplitude $\eta_\mathrm{pr}(t)$ and frequency
$\omega_\mathrm{pr}$. This is described by the driving Hamiltonian
$-\hbar\eta_\mathrm{pr}(t) (\hat{a}e^{i(\delta t + \varphi)}+
\hat{a}^\dag e^{-i(\delta t + \varphi)} )$ with probe-pump detuning
$\delta = \omega_\mathrm{pr}-\omega_\mathrm{p}$ and relative phase
$\varphi$ between probe and pump beam. The resulting coherent
intracavity probe field is given by
\begin{equation}\label{eq:probe-field}\tag{S14}
\alpha_\mathrm{pr}(t) = -\frac{\eta_\mathrm{pr}(t) e^{-i(\delta t
+ \varphi)}}{\tilde{\Delta}_\mathrm{c}+i\kappa}.
\end{equation}
Interference of the probe field with the transverse pump field and
the steady-state cavity field $\alpha_0$ results in a modulated
lattice potential. The corresponding perturbation of the atomic
field is given by
\begin{equation}\label{eq:effective-hamiltonian-probe}\tag{S15}
\hat{H}_\mathrm{pr}=
\hbar\xi(t)\biggl[\delta\hat{\Theta}+\frac{U_0\Theta_0}{\tilde{\Delta}_\mathrm{c}+i\kappa}\delta\hat{\mathcal{B}}\biggl]\cos(\delta
t + \varphi),
\end{equation}
with perturbation amplitude $\xi(t)=2\eta\sqrt{n_\mathrm{pr}(t)}$
and mean intracavity photon number $n_\mathrm{pr}(t)=
\frac{\eta^2_\mathrm{pr}(t)}{\tilde{\Delta}_\mathrm{c}^2+\kappa^2}$.
As $|\tilde{\Delta}_\mathrm{c}| \gg \kappa$ in the experiment, we
set the phase shift, originating from coupling into the cavity, to
$\pi$. In terms of the mode operator $\hat{c}_1$, the perturbation
reads
\begin{equation}\label{eq:effective-probe-two-modes}\tag{S16}
\hat{H}_\mathrm{pr}= \hbar\xi(t)
\sqrt{N\chi}(\hat{c}_1+\hat{c}_1^\dag) \cos(\delta t + \varphi)\,.
\end{equation}
To quantify the response on this perturbation, we evolve the
Hamiltonian $\hat{H} = \hat{H}_\mathrm{exc}+ \hat{H}_\mathrm{pr}$
in time according to the Heisenberg equation
\begin{equation}\label{eq:eom-c1}\tag{S17}
\begin{split}
i\hbar\dot{\hat{c}}_1 = & E_1\hat{c}_1 + 2
NV\chi(\hat{c}_1+\hat{c}_1^\dag) \\
&+ \hbar
\xi(t)\sqrt{N\chi}\cos(\delta t + \varphi)-i\hbar \gamma \hat{c}_1\,.
\end{split}
\end{equation}
Motivated by our experimental observations, we phenomenologically
introduced a damping term with damping constant $\gamma$ into the
time evolution of $\hat{c}_1$. This accounts for possible damping
mechanisms like s-wave scattering with other momentum modes, trap
loss or finite-size dephasing. The general solution of
Eq.~\eqref{eq:eom-c1} reads
\begin{equation}\label{eq:solution-c1}\tag{S18}
\hat{c}_1(t) = 2 \eta \sqrt{N\chi n_\mathrm{pr,0}}\biggl(
\frac{E_1}{E_\mathrm{s}}\mathrm{Im}(\mathcal{Y}(t)) +
i\mathrm{Re}(\mathcal{Y}(t))  \biggl)\,.
\end{equation}
Here,
\begin{equation}\label{eq:response-prefactor}\tag{S19}
\mathcal{Y}(t)=e^{(i\omega_\mathrm{s}-\gamma)t}\int_0^t
\mathrm{d}t'\,e^{-(i\omega_\mathrm{s}-\gamma)t'}\cos(\delta
t'+\varphi)\Pi(t')\,,
\end{equation}
with $\omega_\mathrm{s}=E_\mathrm{s}/\hbar$ and
$n_\mathrm{pr}(t)=n_\mathrm{pr,0}\Pi(t)^2$. Here, the envelope
function $\Pi(t)$ of the probe pulse with duration $\tau$ and the
maximum probe photon number $n_\mathrm{pr,0}$ were introduced. The
population of the excited momentum mode and the mean intracavity
photon number are directly obtained from Eq.~\eqref{eq:solution-c1}
\begin{equation}\label{eq:final-solution}\tag{S20}
\begin{split}
N_1(t) &= \langle \hat{c}_1^\dag \hat{c}_1\rangle  \\
& = 4 \eta^2n_\mathrm{pr,0}\, N\, \chi\,\biggl[ \biggl( \frac{E_1}{E_\mathrm{s}}\biggl)^2\mathrm{Im}(\mathcal{Y}(t))^2 + \mathrm{Re}(\mathcal{Y}(t))^2  \biggl]\\
n_\mathrm{ph}(t)& = \bigg|\alpha_0 - \frac{4\eta^2
\sqrt{n_\mathrm{pr,0}}
N\chi}{\tilde{\Delta}_\mathrm{c}+i\kappa}\biggl(
\frac{E_1}{E_\mathrm{s}}\biggl)\mathrm{Im}(\mathcal{Y}(t))
\end{split}
\end{equation}
\\[-43pt]
\begin{equation}\label{eq:final-solution}\notag
\begin{split}
&+\sqrt{n_\mathrm{pr}(t)}e^{-i(\delta t + \varphi)}\bigg|^2\,.
\end{split}
\end{equation}
The second term in  $n_\mathrm{ph}(t)$ describes the pump field
which was Bragg scattered off the excited density modulation into
the cavity mode. From $N_1$, the population $N_\mathrm{e}$ in the
momentum state $|\mathrm{e}\rangle$ is obtained according to
$N_\mathrm{e} = \zeta N_1$, where $\zeta$ denotes the absolute
square of the Fourier amplitude of the excited Bogoliubov mode
$u_1 + v_1$ at momenta $(\pm\hbar k, \pm\hbar k)$.

The curves shown in Fig.~2C and D of the main text are obtained
from Eqs.~\eqref{eq:final-solution}, where the phase $\varphi$ was
adjusted to fit the data in Fig.~2C. The resonance curves,
displayed in Fig. 2B and E of the main text, correspond to
phase-averaged values $\langle N_\mathrm{e}(\tau)
\rangle_{\varphi\in [0,2\pi]}$ and $\langle \int_0^\tau
\,\mathrm{d}t \,n_\mathrm{ph}(t) /\tau\rangle_{\varphi\in
[0,2\pi]}$ where the resonance frequency $\omega_\mathrm{s}$ and
the amplitude $\eta^2 N \chi$ were adjusted independently to fit
the data. Due to the detection background an offset was added to
$N_\mathrm{e}(\tau)$. The standard deviation of fluctuations
associated with the uncontrolled relative phase $\varphi$ is
displayed by the shadings in Fig.~2B and E of the main text.

\subsection*{Response in atomic density and momentum state population}
\emph{Bragg spectroscopy, as described by the density perturbation
$\hat{H}_\mathrm{pr}$, induces a response in the atomic density
and in the population of the excited state. Normalized to the
integrated amplitude of the perturbation, this density response is
a measure for the susceptibility of the system on an external
density perturbation and provides at zero temperature a direct
link to the static structure factor
\cite{pitaevskii2003,ozeri2005}. We use the recorded modulation
$\langle\delta\hat{a}\rangle$ of the intracavity field to quantify
the density response $\langle \delta \hat{\rho}\rangle$ of the
system.}

Consider the density operator
$\hat{\rho}(\mathbf{r})=\hat{\Psi}^\dag(\mathbf{r})\hat{\Psi}(\mathbf{r})$
 and its linear expansion $\hat{\rho}= \rho_0 +
\delta\hat{\rho}=N|\psi_0|^2 +
\sqrt{N}\psi_0(\delta\hat{\Psi}^\dag+\delta\hat{\Psi})$ around the
equilibrium value $\rho_0=\langle \hat{\rho}\rangle$. In terms of
the density fluctuation operator in Fourier space, defined as
$\delta\hat{\rho}_\mathbf{k}=\int \mathrm{d}^3r\,
e^{-i\mathbf{k}\mathbf{r}}\delta\hat{\rho}(\mathbf{r})$, the
fluctuations of the order and bunching parameter
$\delta\hat{\Theta}$ and $\delta\hat{\mathcal{B}}$ read
\begin{equation}\label{eq:density-component-k1}\notag
\delta\hat{\Theta} = \sum_{\mathbf{k}\in (\pm k, \pm k)}
\delta\hat{\rho}_{\mathbf{k}}/4  \qquad \text{and} \qquad
\delta\hat{\mathcal{B}} = \sum_{\mathbf{k}\in (\pm 2 k, 0)}
\delta\hat{\rho}_{\mathbf{k}}/4\,.
\end{equation}
Fluctuations of the cavity field $\delta \hat{a}$, see
Eq.~\eqref{eq:delta-a}, are thus given by
\begin{equation}\label{eq:field-fluctuations}\notag
\delta \hat{a} = \frac{\eta}{4(\tilde{\Delta}_\mathrm{c}+i\kappa)}
\left[\sum_{\mathbf{k}\in (\pm k, \pm k)}
\delta\hat{\rho}_{\mathbf{k}}  + \frac{\Theta_0
U_0}{\tilde{\Delta}_\mathrm{c}+i\kappa} \sum_{\mathbf{k}\in (\pm 2
k, 0)} \delta\hat{\rho}_{\mathbf{k}} \right]\,,
\end{equation}
and provide a measure of the induced atomic density modulation. We
define the corresponding (quadratic) density response function
\begin{equation}\label{eq:density-response}\notag
\small{
%\mathcal{R}_{(\delta\hat{\rho}_{\mathbf{k}_1},\delta\hat{\rho}_{\mathbf{k}_2})}  =
\mathcal{R}_{\rho}  = \frac{1}{P \int_0^\tau n_\mathrm{pr}(t)dt}
\int_{-\infty}^\infty d\delta \left\langle\int_0^\tau 
\frac{|\langle\delta
\hat{a}\rangle|^2 dt}{|{\eta/(4(\tilde{\Delta}_\mathrm{c}+i\kappa))}|^2}\right
\rangle _{\varphi \in [0, 2\pi]}\,,
}
\end{equation}
where again an average over the relative phase $\varphi$ is
performed.

Experimentally, we extract $|\langle \delta \hat{a}\rangle|^2$
from the resonance fits based on Eq.~\eqref{eq:final-solution}
(see Fig.~2E in the main text), with experimentally calibrated
$|\alpha_0|^2$ and $n_\mathrm{pr}$. In order to probe the linear
response of the system, the probe power was accordingly lowered
when approaching the critical point. The lowest perturbation
applied in the normal phase corresponds to $30(8)$ intracavity
probe photons in total.

Similarly, the response function $\mathcal{R}_{N}$ associated with
the detected number of atoms with momenta $(\pm \hbar k, \pm \hbar
k)$ is defined as
\begin{equation}\label{eq:atom-response-theory}\notag
\mathcal{R}_{N} = \frac{1}{P \int_0^\tau \mathrm{d}t\,
n_\mathrm{pr}(t)} \int_{-\infty}^\infty \mathrm{d}\delta\, \langle
N_\mathrm{e}(\tau) \rangle _{\varphi \in [0, 2\pi]}.
\end{equation}
which is proportional to the area below the fitted resonances of
$N_\mathrm{e}(\tau)$, see Fig.~2B in the main text.

The dominant scaling factor of $\mathcal{R}_\rho$ when approaching
the critical point is found from Eq.~\eqref{eq:final-solution} to
be $(E_1/E_\mathrm{s})^2$. As $\mathcal{R}_\rho$ measures the
quadratic density response, this is in agreement with Feynman's
relation, given in the main text. Since $N_\mathrm{e}(\tau)$ is
sensitive to both quadratures of $\hat{c}_1$ (see
Eq.~\eqref{eq:final-solution}), the response function
$\mathcal{R}_{N}$ scales differently and exhibits larger variances
in the relative phase $\varphi$.

\end{appendix}


\begin{thebibliography}{18}%



\bibitem{bloch2008}
I.~{B}loch, J.~{D}alibard, W.~{Z}werger, {\it {R}ev. {M}od. {P}hys.\/} {\bf 80}, 885 (2008).

\bibitem{griesmaier2005}
A.~{G}riesmaier, J.~{W}erner, S.~{H}ensler, J.~{S}tuhler,
T.~{P}fau, {\it {P}hys. {R}ev.
  {L}ett.\/} {\bf 94}, 160401 (2005).

\bibitem{ni2008}
K.-K. {N}i, {\it et~al.\/}, {\it {S}cience\/} {\bf 322}, 231 (2008).

\bibitem{deiglmayr2008}
J.~{D}eiglmayr, {\it et~al.\/}, {\it {P}hys. {R}ev.
  {L}ett.\/} {\bf 101}, 133004 (2008).

\bibitem{lu2011}
M.~{L}u, N.~Q. {B}urdick, S.~H. {Y}oun, B.~L. {L}ev, {\it {P}hys. {R}ev.
  {L}ett.\/} {\bf 107}, 190401 (2011).

\bibitem{lahaye2009}
T.~{L}ahaye, C.~{M}enotti, L.~{S}antos, M.~{L}ewenstein,
T.~{P}fau, {\it {R}ep. {P}rog. {P}hys.\/}
  {\bf 72}, 126401 (2009).

\bibitem{santos2003}
L.~{S}antos, G.~V. {S}hlyapnikov, M.~{L}ewenstein, {\it {P}hys. {R}ev. {L}ett.\/} {\bf 90}, 250403 (2003).

\bibitem{o'dell2003}
D.~H.~J. {O}'{D}ell, S.~{G}iovanazzi, G.~{K}urizki, {\it {P}hys.
  {R}ev. {L}ett.\/} {\bf 90}, 110402 (2003).

\bibitem{cherng2009}
R.~W. {C}herng, E.~{D}emler, {\it {P}hys. {R}ev.
  {L}ett.\/} {\bf 103}, 185301 (2009).

\bibitem{henkel2010}
N.~{H}enkel, R.~{N}ath, T.~{P}ohl, {\it {P}hys. {R}ev. {L}ett.\/} {\bf 104},
  195302 (2010).

\bibitem{yarnell1958}
J.~L. {Y}arnell, G.~P. {A}rnold, P.~J. {B}endt, E.~C. {K}err, {\it
  {P}hys. {R}ev. {L}ett.\/} {\bf 1}, 9 (1958).

\bibitem{pomeau1994}
Y.~{P}omeau, S.~{R}ica, {\it
  {P}hys. {R}ev. {L}ett.\/} {\bf 72}, 2426 (1994).

\bibitem{lahaye2007}
T.~{L}ahaye, {\it et~al.\/}, {\it {N}ature\/} {\bf 448}, 672 (2007).

\bibitem{fattori2008}
M.~{F}attori, {\it et~al.\/}, {\it {P}hys.
  {R}ev. {L}ett.\/} {\bf 101}, 190405 (2008).

\bibitem{bismut2010}
G.~{B}ismut, {\it et~al.\/}, {\it {P}hys. {R}ev. {L}ett.\/} {\bf 105},
  040404 (2010).

\bibitem{koch2008}
T.~{K}och, {\it et~al.\/}, {\it {N}ature {P}hys.\/} {\bf 4}, 218 (2008).

\bibitem{munstermann2000}
P.~{M}\"unstermann, T.~{F}ischer, P.~{M}aunz, P.~W.~H. {P}inkse,
G.~{R}empe,{\it {P}hys. {R}ev. {L}ett.\/} {\bf
  84}, 4068 (2000).

\bibitem{asboth2004}
J.~K. {A}sb\'oth, P.~{D}omokos, H.~{R}itsch, {\it {P}hys. {R}ev. {A}\/}
  {\bf 70}, 013414 (2004).

\bibitem{maschler2005}
C.~{M}aschler, H.~{R}itsch, {\it {P}hys. {R}ev. {L}ett.\/} {\bf 95},
  260401 (2005).

\bibitem{slama2008a}
S.~{S}lama, G.~{K}renz, S.~{B}ux, C.~{Z}immermann, P.~W.
{C}ourteille, {\it {AIP} {C}onf. {P}roc.\/} {\bf
  970}, 319 (2008).

\bibitem{domokos2002}
P.~{D}omokos, H.~{R}itsch, {\it {P}hys. {R}ev.
  {L}ett.\/} {\bf 89}, 253003 (2002).

\bibitem{nagy2008}
D.~{N}agy, G.~{S}zirmai, P.~{D}omokos, {\it {E}ur. {P}hys.
  {J}. {D}\/} {\bf 48}, 127 (2008).

\bibitem{baumann2010}
K.~{B}aumann, C.~{G}uerlin, F.~{B}rennecke, T.~{E}sslinger, {\it
  {N}ature\/} {\bf 464}, 1301 (2010).

\bibitem{stenger1999}
J.~{S}tenger, {\it et~al.\/}, {\it {P}hys. {R}ev. {L}ett.\/} {\bf 82},
  4569 (1999).

\bibitem{SOM}
See supporting material on Science Online.

\bibitem{steinhauer2002}
J.~{S}teinhauer, R.~{O}zeri, N.~{K}atz, N.~{D}avidson, {\it {P}hys. {R}ev.
  {L}ett.\/} {\bf 88}, 120407 (2002).

\bibitem{nozieres1990}
P.~{N}ozi\`eres, D.~{P}ines, {\it {T}he {T}heory of {Q}uantum {L}iquids\/},
  vol.~II (Addison-Wesley, Reading, MA, 1990).


\bibitem{gopalakrishnan2009}
S.~{G}opalakrishnan, B.~L. {L}ev, P.~M. {G}oldbart, {\it {N}ature {P}hys.\/} {\bf 5}, 845 (2009).

\bibitem{strack2011}
P.~{S}track, S.~{S}achdev, {\it {P}hys. {R}ev. {L}ett.\/} {\bf 107}, 277202 (2011).

\bibitem{gopalakrishnan2011}
S.~{G}opalakrishnan, B.~L. {L}ev, P.~M. {G}oldbart,
  {\it {P}hys. {R}ev. {L}ett.\/} {\bf 107}, 277201 (2011).



\end{thebibliography}

\begin{thebibliography}{40}%
%\bibpunct{(S}{)}{;}{a}{,}{,}

\bibitem[31]{morsch2006}
O.~{M}orsch, M.~{O}berthaler, {\it {R}ev. {M}od. {P}hys.\/} {\bf
78}, 179
  (2006).

\bibitem[32]{baumann2010}
K.~{B}aumann, C.~{G}uerlin, F.~{B}rennecke, T.~{E}sslinger, {\it
{N}ature\/}
  {\bf 464}, 1301 (2010).

\bibitem[33]{maschler2005}
C.~{M}aschler, H.~{R}itsch, {\it {P}hys. {R}ev. {L}ett.\/} {\bf
95}, 260401
  (2005).

\bibitem[34]{maschler2008}
C.~{M}aschler, I.~B. {M}ekhov, H.~{R}itsch, {\it {E}ur. {P}hys.
{J}. {D}\/}
  {\bf 46}, 545 (2008).

\bibitem[35]{nagy2010}
D.~{N}agy, G.~{K}onya, G.~{S}zirmai, P.~{D}omokos, {\it {P}hys.
{R}ev.
  {L}ett.\/} {\bf 104}, 130401 (2010).

\bibitem[36]{nagy2008}
D.~{N}agy, G.~{S}zirmai, P.~{D}omokos, {\it {E}ur. {P}hys. {J}.
{D}\/} {\bf
  48}, 127 (2008).

\bibitem[37]{kraemer2003}
M.~{K}r\"amer, C.~{M}enotti, L.~{P}itaevskii, S.~{S}tringari, {\it
{T}he
  {E}uropean {P}hysical {J}ournal {D} - {A}tomic, {M}olecular, {O}ptical and
  {P}lasma {P}hysics\/} {\bf 27}, 247 (2003).

\bibitem[38]{pitaevskii2003}
L.~{P}itaevskii, S.~{S}tringari, {\it {B}ose-{E}instein
{C}ondensation\/}
  (Oxford University Press, 2003).

\bibitem[39]{castin2001}
Y.~{C}astin, {\it {C}oherent atomic matter waves\/} (Springer
Berlin, 2001),
  vol.~72, chap. Course 1: Bose-Einstein Condensates in Atomic Gases: Simple
  Theoretical Results, pp. 1--136.

\bibitem[40]{ozeri2005}
R.~{O}zeri, N.~{K}atz, J.~{S}teinhauer, N.~{D}avidson, {\it {R}ev.
{M}od.
  {P}hys.\/} {\bf 77}, 187 (2005).



\end{thebibliography}
\end{document}